\documentclass[12pt,sort&compress]{article}

\usepackage[dvips]{color}
\usepackage{amsmath}
\usepackage{graphicx}
\usepackage{tabularx}
\usepackage{subfig}
\usepackage{cite}
\usepackage{amsfonts}
\usepackage{tabularx,ragged2e}
\usepackage{tikz}
\usepackage{tkz-tab}
\usepackage{xcolor}
\newcolumntype{C}{>{\Centering\arraybackslash}X} 
\usepackage[outercaption]{sidecap}    
\usepackage{hyperref}

\newcommand{\rhohat}{\hat{\rho}}

\newcommand{\rmin}{r_-}
\newcommand{\rmax}{r_+}

\begin{document} 
\title{Renewal processes with a trap  under stochastic resetting}
\date{}
\author{Pascal Grange\\
{\emph{Division of Natural and Applied Sciences}}\\
{\emph{and Zu Chongzhi Center for Mathematics and Computational Science}}\\
 Duke Kunshan University\\
8 Duke Avenue, Kunshan, 215316 Jiangsu, China\\
\normalsize{{\ttfamily{pascal.grange@dukekunshan.edu.cn}}}}
\maketitle

\begin{abstract}
 Renewal processes are zero-dimensional processes defined by independent intervals of time between 
 zero crossings of a random walker. We subject renewal processes them to stochastic resetting by setting the position of 
 the random walker to the origin at Poisson-distributed time with rate $r$. We introduce an additional parameter, the probability $\beta$
 of keeping the sign state of the system  at resetting time. Moreover, we introduce a trap at the origin, which absorbs the process
 with a fixed probability at each zero crossing. We obtain the mean lifetime of the process in closed form.  For time intervals  
 drawn from a L\'evy stable distribution of parameter $\theta$, the mean lifetime is finite for every positive value of the resetting rate, 
 but goes to infinity when $r$ goes to zero. If the sign-keeping probability
 $\beta$ is higher than a critical level $\beta_c(\theta)$ (and strictly lower than $1$), the mean lifetime exhibits two extrema as a function of the resetting rate.  Moreover, it goes to zero as $r^{-1}$ when $r$ goes to infinity. On the other hand, there is a single minimum if $\beta$ is set to one.

\end{abstract}

\tableofcontents

\section{Introduction}

Stochastic resetting puts the system in contact with initial conditions at Poisson-distributed times, which can lead to new 
   out-of-equilibrium steady states. Resetting events cut off long excursions, which can make some 
    expectation values finite. The first example of such a behavior was discovered in the case of a Brownian 
     random walker in one dimension,  whose mean first-passage time at a fixed target was calculated as a 
      function of the resetting rate, and found to exhibit a minimum \cite{evans2011diffusion,evans2011optimal}.  
       Stochastic resetting has yielded new stationary states and exact results on observables of a variety of  stochastic processes 
 and out-of-equlibrium physical systems (including population dynamics, reaction-diffusion systems and active particles)   \cite{gupta2014fluctuating,evans2018run,refractory,ZRPSS,ZRPResetting,grange2020entropy,durang2014statistical,magoni2020ising,grange2021aggregation,grange2020susceptibility,perfetto2021designing,toledo2022first,grange2022winding,sarkar2022synchronization,maso2022conditioned,grangeVoterModel}.\\

  In this work we  subject renewal processes to stochastic resetting, and address their lifetime in the presence of a trap 
 using a renewal argument. Renewal processes are zero-dimensional stochastic processes,  which serve as exactly solvable models
 of occupation times. Occupations times  are of crucial interest in the problem of phase persistence in physical systems with interacting degrees of freedom include coalescing droplets and quenched spin systems \cite{beysens1986growth,bray1994non,derrida1996exact,majumdar1996nontrivial}.
 Renewal processes  have been extensively studied in \cite{baldassarri1999statistics,godreche2001statistics,de2001statistics,lamperti,godreche2015statistics,godreche2019non} (see also \cite{godreche2023poisson} for developments on the ordering of observables of renewal processes subjected to stochastic resetting). They are generalizations of Poisson processes to cases where the intervals between events are not exponentially distributed, and are defined as follows. 
  Consider the sign  of a real-valued stochastic process, which we can think of as the position of a random walker starting at the origin and 
 drawing  the increments of its position from some fixed distribution. Let us  denote by $\rho$  the
 fixed density of the next return time to the origin of the random walker.
    The walker crosses the origin  at random times called epochs. By convention there is an epoch at $t_0=0$. At each epoch $t_n$, for $n\geq 0$, a random interval  
   $\tau_{n+1}$ is  drawn from  the  distribution $\rho$, and the next  epoch is defined as $t_{n+1} = t_n + \tau_{n+1}$.  
    Moreover, let us imagine that a random sign $\sigma=\pm1$,  is drawn at $t_0$. We say that the system spends the 
    interval $[0,t_1[$ in the sign state $\sigma$, and the sign changes at each epoch.
   With these definitions, $t_n = \tau_1 + \dots + \tau_n$, and the system is in the state $\sigma$ (resp. 
    $-\sigma$)  in each interval of the form $[t_{2k}, t_{2k+1}[$ (resp. $[t_{2k+1}, t_{2k+2}[$) for  $k\in \mathbb{N}$. 
    The 
     distribution on the number of sign switches in $[0,t]$, backward and forward time to the nearest  sign switch, and total time spent in 
      a given sign state  in $[0,t]$, have been studied in \cite{godreche2001statistics}. The behavior of these observable was found to be dependent 
       on whether the distribution $\rho$ is narrow (i.e. possesses moments of every order) or broad.\\
       
      Let us subject the process to stochastic resetting. At Poisson-distributed times (with rate $r>0$) the position of the  random walker is set at the origin, and a new time is drawn 
 from the distribution $\rho$. We have to define on which side of the origin the walker goes immediately after this resetting event: we will 
 let the walker keep its current sign state with probability $\beta$ in $]0,1[$ (and let it switch its sign state with probability $(1-\beta)$).   Moreover, let us introduce a trap at the origin: every time the process crosses the origin, it survives with some fixed probability $\alpha<1$. Once the density $\rho$ has been chosen, the process has three parameters: the resetting rate $r$, the persistence probability $\beta$ and the survival probability $\alpha$. \\

 The average lifetime of the process is intuitively an increasing function of the survival rate $\alpha$, but we may wonder how it depends on the resetting rate $r$ (for phase diagrams of stochatic processes with stochastic resetting driven by broad distributions, see \cite{kusmierz2014first,campos2015phase}). If the sign-keeping probability $\beta$ is large enough, and if the distribution $\rho$ is broad enough, it could be beneficial to increase the resetting $r$. Indeed, one may expect that more frequent resetting events increase the probability to draw a large time interval. 


The paper is organized as follows. 
 In Section \ref{QOI} we review the definition of renewal processes and define the resetting prescription. 
 In Section \ref{mean} we  calculate the mean lifetime of the process using renewal equations (renewal equations are powerful tools to estimate observables of systems subjected to stochastic resetting, see for instance \cite{evans2018run,refractory,maso2022conditioned}). In Section \ref{extrema} 
 study the variations of this mean lifetime for a process driven by a positive L\'evy stable distribution of index $\theta$ 
 and obtain the phase diagram of the model in the $(\beta,\theta)$ square. It is characterized by a critical level of the  sign-keeping probability, denoted by $\beta_c(\theta)$ (above which two extrema of the mean lifetime exist). We obtain the graph of $\beta_c$  from parametric equations, and derive local expressions of 
$\beta_c$ for $\theta$ close to $0$ or $1$. 



\section{Definitions, notations and quantities of interest}\label{QOI}       
    
  Consider a random walker on the real line, in continuous time,  and denote by $x_t$ its position 
   at time $t$. The position is initialized at the origin, $x_0=0$.
   Every time the walker is at the 
    origin, a random variable is drawn from a fixed probability density $\rho$, which defines the time to the next zero crossing.
  Such a process is called a renewal process (and each zero crossing is called a renewal).  In a renewal process, the times between zero crossings are independent,
       identically distributed (i.i.d.) variables, denoted by $(\tau_n)_{n\geq 1}$. By convention $t=0$ is considered a zero
        crossing, hence we can express the time $t_n$ at which the $n$-th zero crossing occurs:
        \begin{equation}
        \begin{split}
       t_0 &= 0, \\
       t_n &= \sum_{k=1}^n \tau_k,\;\;\;\;\;\;\;\;\;\;\;(n\geq 1).
       \end{split}
        \end{equation}
 The sign state of the process is defined as
\begin{equation}
 \sigma_t = {\mathrm{sign}}( x_t).
\end{equation}
     The process starts at time $0$, with $x_0=0$. At time $0$ the initial sign state of the position (positive or negative) 
    is drawn uniformly from $\{-1,+1  \}$. If this initial sign state is positive (resp. negative), the walker spends the time $\tau_1$ in a positive (resp. negative) sign state.
 The sign state will change at $\tau_1$, when $\tau_2$ is drawn. With these rules the sign state is defined at all positive times. 
  If $\rho$ is chosen to be a positive stable L\'evy distribution of index $\theta$ in $]0,1[$, the sign state corresponds to model the magnetization of a single spin with a broad distribution 
 of independent time intervals between flips introduced in \cite{baldassarri1999statistics}  (generalization to higher values of $\theta$ and other observables appeared in \cite{godreche2001statistics}).\\

     Let us subject the process to stochastic resetting.  In an infinitesimal interval $[t, t + dt]$, the position of the walker is reset to the origin with probability 
       $rdt$, where $r>0$ is a fixed resetting rate.    At such  a resetting event, a new time is drawn from the distribution $\rho$.
 Moreover,
       the current sign  state is kept with a fixed probability $\beta$ (called the sign-keeping probability) in the interval $]0,1[$, and switched with 
        probability $1-\beta$. If the sign is switched, the resetting event is also a zero crossing. There are therefore 
         two kinds of zero crossings:\\
       1. renewal events (which occur if no resetting has happened before the end of the latest drawn random time),\\
       2. resetting events at which the sign state is switched.\\

    Let us introduce a trap at the origin: at each zero crossing the walker survives with probability $\alpha$. With  this convention, the random walker survives  a resetting event with probability $\beta+ \alpha(1-\beta)$. Indeed, given that the sign state is kept at the resetting event, there is no zero crossing. 
 Consider the lifetime $T^{(\alpha,r,\beta)}$ of the process. Its mean is denoted by
  $\left\langle  T^{(\alpha,r,\beta)}\right\rangle$. The brackets denote averages over all the involved random processes (renewal times, resetting times with their random sign, absorption by the trap).\\

 At time $t>0$, the process  has survived with a probability denoted by $Q^{(\alpha,r,\beta)}(t)$.
 The  death rate of the process  is the opposite of the  derivative of this survival probability. 
  The  mean lifetime of the process is given by integating time against this death rate:
\begin{equation}
\left\langle  T^{(\alpha,r,\beta)}\right\rangle = - \int_0^\infty t \frac{dQ^{(\alpha,r,\beta)}}{dt}(t) dt. 
\end{equation} 
 Integrating by parts yields  the mean lifetime of the process as the value at zero of the Laplace transform 
 of the survival probability:
\begin{equation}\label{Trb}
 \left\langle  T^{(\alpha,r,\beta)}\right\rangle = \hat{Q}^{(\alpha,r,\beta)}(0),\\
\end{equation}
 where $\hat{F}$ denotes  the Laplace transform with respect to time of any function $F$:
\begin{equation}\label{LaplaceDef}
 \hat{F}( s) := \int_0^\infty e^{-st} F(t) dt.
\end{equation} 
 We are therefore instructed to work out the Laplace transform of the survival probability  $Q^{(\alpha,r,\beta)}$ of the process.\\

 In the next section we will  condition on the latest resetting time,   to obtain a renewal equation for  the survival probability.
 We will naturally be led to consider the number $N_t$ of sign switches in the interval $[0,t]$, whose probability law depends on the resetting rate $r$ and on the  probabilities $\alpha$ and $\beta$. Let us denote by 
$p_n^{(\alpha,r,\beta)}(t)$ the probability of the value $n$:
\begin{equation}
    p_n^{(\alpha,r,\beta)}(t):={\mathrm{Prob}}\left( N_t =n\right) ,\;\;\;\;\; (n\geq0).
\end{equation}
 The corresponding probabilities are denoted by $p_n$ in the process without resetting (meaning $r=0$) and without trap (meaning $\alpha=1$): 
\begin{equation}\label{pnnot}
    p_n(t):= p_n^{(\alpha = 1,r=0,\beta)}(t).
\end{equation}
 When the resetting rate is zero, the parameter $\beta$ plays no role because it is only involved in resetting events.
 The notation $p_n$ matches the one used in \cite{godreche2001statistics}, where the Laplace transform $\hat{p}_n$
 has been worked out  in terms of the Laplace transform $\hat{\rho}$ of the probability density 
 of times between renewals:
\begin{equation}\label{pn0}
  \hat{p}^{(0)}_n(s) = \rhohat(s)^n\frac{1-\rhohat(s)}{s},\;\;\;\;\;\;\;\;\;(n\geq 0 ).
 \end{equation}
 Let us review the derivation. With the convention $t_0=0$, the probability $p_n^{(0)}(t)$ is the expected value 
 of the random variable $\mathbf{1}( t_n < t < t_{n+1})$, where  $\mathbf{1}$ denotes the indicator function. For $n\geq 0$, the Laplace transform of this identity yields 
   \begin{equation}
   \begin{split}
    \hat{p}^{(0)}_n(s) &= \int_0^\infty dt e^{-st} p_n^{(0)}(t) dt
     =   \left\langle  \int_0^\infty dt e^{-st}\mathbf{1}( t_n < t < t_{n+1}) dt  \right\rangle 
        =   \left\langle  \int_{t_n}^{t_{n+1}}  e^{-st}  dt  \right\rangle \\
     & =   \frac{1}{s}\langle  e^{-s t_n} -  e^{-s t_{n+1}}\rangle
      =     \frac{1}{s}\langle  e^{-s(\tau_1+\dots+ \tau_n)}( 1 -  e^{-s \tau_{n+1}})\rangle,
   \end{split}
   \end{equation}
   which yields Eq. (\ref{pn0}) using the independence of the random variables $(\tau_1,\dots,\tau_{n+1})$.\\


\section{Renewal equation for the survival probability}\label{mean}
  
  Consider a configuration at time $t>0$ in which the process has survived.  
Let us condition on the latest resetting time (if any) in $[0,t]$, call it $t-\tau$. The process has survived 
 this resetting event, which has happened  in one of the  following two ways:\\
1. the random walker has  not crossed the origin at $t-\tau$ (probability $\beta$),\\
2.  the random walker has crossed the origin at $t-\tau$ and has survived this crossing (probability $(1-\beta)\alpha$).\\
 After this latest resetting, the process has evolved as a renewal process without resetting, with a trap at the origin.
 It has undergone  $n$ renewals (for some $n\geq 0$) between $t-\tau$ and $t$, and has survived each of them.
 The probability of such an event in $]t-\tau,t]$ is $\alpha^n p_n(\tau)$, in the notation introduced in  Eq. (\ref{pnnot}).\\

 Let us denote by the $Q^{(\alpha,0)}(t)$ the survival probability at time $t$ of process without resetting (if there is no resetting the parameter $\beta$ plays no role, and the zero in the notation refers to $r=0$).
Moreover, the probability survival probability at time $t$ without any resetting event in $[0,t]$ equals 
$e^{-rt}Q^{(\alpha,0)}(t)$.
 Conditioning on the latest resetting event therefore induces the following renewal equation:
\begin{equation}\label{renoQ}
\begin{split}
Q^{(\alpha, r,\beta)}(t) =& e^{-rt}Q^{(\alpha,0)}(t) + r\beta \int_0^t d\tau e^{-r\tau}  Q^{(\alpha,r,\beta)}(t-\tau) \sum_{n\geq 0} p^{(0)}_n(\tau)\alpha^n\\
      &+ r(1-\beta) \alpha  \int_0^t d\tau e^{-r\tau}  Q^{(\alpha,r,\beta)}(t-\tau) \sum_{n\geq 0} p^{(0)}_n(\tau)\alpha^n\\
=&  e^{-rt}Q^{(0)}(t) + r\beta \int_0^t d\tau e^{-r\tau}  Q^{(\alpha,r,\beta)}(t-\tau) G^{(0)}(\tau,\alpha) \\
  &+ r(1-\beta) \alpha  \int_0^t d\tau e^{-r\tau}  Q^{(\alpha,r,\beta)}(t-\tau) G^{(0)}(\tau,\alpha),
\end{split}
\end{equation}
where we have introduced the notation  $G^{(0)}$ for the generating function of the probability law of the number of sign switches in the process without resetting:
\begin{equation}
G^{(0)}(t,x) :=  \sum_{n\geq 0} p_n(t) x^n.
\end{equation}

The Laplace transform w.r.t. the time variable (denoted by a hat as in Eq. (\ref{LaplaceDef})) maps convolutions to ordinary products. The Laplace transform of Eq. (\ref{renoQ}) therefore reads
\begin{equation}
\hat{Q}^{(r,\beta)}(s) = \hat{Q}^{(0)}(r+s) + r\beta  \hat{Q}^{(r,\beta)}(s)  \hat{G}^{(0)}(r+s,\alpha)
 + r(1-\beta) \alpha \hat{Q}^{(r,\beta)}(s)   \hat{G}^{(0)}(r+s,\alpha).  
\end{equation}
  Substituting $0$ to $s$ and using Eq. (\ref{Trb}) we obtain the mean lifetime of the process as
\begin{equation}\label{TrExpr}
  \left\langle  T^{(\alpha, r,\beta)}\right\rangle = \frac{\hat{G}^{(0)}(r,\alpha)}{1 -r(\beta + \alpha(1-\beta) ) \hat{G}^{(0)}(r,\alpha)}.
\end{equation}

Moreover, the needed generating function $\hat{G}^{(0)}$ is obtained from Eq. (\ref{pn0}) as a geometric sum:
\begin{equation}
 \hat{G}^{(0)}(s,\alpha) = \frac{ 1-\rhohat(s)}{  s(  1-\alpha\rhohat(s) ) }.
\end{equation}
Substituting into Eq. (\ref{TrExpr}) yields
\begin{equation}\label{TrExprF}
\begin{split}
  \left\langle  T^{(\alpha,r,\beta)}\right\rangle =& \frac{    1-\rhohat(r)   }{   r( 1-\alpha\rhohat(r)  )     -r(\beta + \alpha(1-\beta) ) ( 1-\rhohat(r) )      }\\
 =&   \frac{    1-\rhohat(r)   }{   r [ 1-\alpha - \beta +\alpha\beta +\beta(1-\alpha) \rhohat( r ) ]      }\\
=&   \frac{    1-\rhohat(r)   }{   (1-\alpha)r [ 1- \beta  +\beta \rhohat( r ) ]      }.
\end{split}
\end{equation}
 The dependence on  the survival rate $\alpha$ comes from a factor that does depend only  on $\alpha$. The existence  of extrema of the mean lifetime of the process in the resetting rate $r$ therefore depends only on the parameter $\beta$ and on the distribution. 

\section{The mean lifetime for a broad distribution of renewal times}\label{extrema}
\subsection{Existence of extrema of the mean lifetime}

 Consider a renewal process driven by a positive stable L\'evy deistribution, whose Laplace transform\footnote{we could have considered $\rhohat(s) = \exp( -a s^\theta)$ for some fixed parameter $a$. Setting $a$ to $1$ is equivalent to choosing the time scale of the process.} is given explicitly \cite{pollard1946representation,penson2010exact} by 
 \begin{equation}
 \rhohat(s) = \exp( - s^\theta),\;\;\;\;\;{\mathrm{with}}\;\;\;\;\;0<\theta<1.
\end{equation}
 The distribution $\rho$ of time intervals between renewals is broad, in the sense that it does not have a finite first moment, because
 \begin{equation}
 \rhohat(s) \underset{s\to 0}{=}  1 - s^\theta + o(s^{\theta}).
\end{equation}
 
From the expression obtained  in Eq. (\ref{TrExprF}), we observe that the mean lifetime goes to zero 
 at large resetting rate:
\begin{equation}
 \left\langle  T^{(\alpha,r,\beta)}\right\rangle  \underset{r\gg 1}{\sim} \frac{1}{(1-\alpha)(1-\beta)r},
\end{equation}
 which is intuitive because a larger resetting rate results in more  frequent exposures of  the process to the trap. On the other hand,
  the mean lifetime goes to infinity when the resetting rate goes to zero:
\begin{equation}
 \left\langle  T^{(\alpha,r,\beta)}\right\rangle  \underset{r\to 0}{\sim} \frac{r^{\theta-1}}{1-\alpha},
\end{equation}
 This property is a consequence of the broad nature of the distribution $\rho$, and the above equivalent
 does not depend on the sign-keeping probability $\beta$. In between these two regimes, the lifetime may be
 a decreasing function of the resetting rate, or exhibit local extrema, depending on the values of  the 
 parameters $\beta$ and $\theta$.\\

Substituting the chosen expression of $\hat{\rho}$ into the expression obtained in Eq. (\ref{TrExprF}) yields the mean lifetime of the process as
  \begin{equation}
\left\langle  T^{(\alpha, r,\beta)}\right\rangle = \frac{1}{(1-\alpha)(1-\beta)}\frac{    1-\exp( - r^\theta)  }{   r  [ 1+ b \exp( -r^\theta) ]      }, \;\;\;\;\;\;\;{\mathrm{with}}\;\;\;\;\;\;\;b:=\frac{\beta}{1-\beta}.
\end{equation}
 The parameter $b$, which is the ratio
 of the persistence to the non-persistence of the sign state at each resetting, can take any positive value.\\

 The logarithmic derivative of the expected lifetime w.r.t. the resetting rate reads
\begin{equation}
\begin{split}
 \frac{\partial}{\partial r}\left( \log \left\langle  T^{(r,\beta)}\right\rangle \right)= & \frac{\theta r^{\theta - 1} e^{-r^\theta}}{ 1- e^{-r^\theta}} - \frac{1}{r} 
            +\frac{ b\theta r^{\theta - 1} e^{-r^\theta} }{ 1 + b e^{-r^\theta}}\\
 =& \frac{ r(1+ b e^{-r^\theta})\theta r^{\theta - 1} e^{-r^\theta}  -(1- e^{-r^\theta})(1 + b e^{-r^\theta}) + r(1- e^{-r^\theta})b\theta r^{\theta - 1}  e^{-r^\theta}  }{r(  1- e^{-r^\theta})  ( 1 + b e^{-r^\theta})}\\
=&   \frac{ (1+ b  ) \theta r^\theta e^{-r^\theta}  -(1- e^{-r^\theta})(1 + b e^{-r^\theta})} {r(  1- e^{-r^\theta})  ( 1 + b e^{-r^\theta})}.\\
\end{split}
 \end{equation}
The extrema of the average lifetime therefore satisfy
\begin{equation}\label{extremaEq}
 r^{\theta} e^{-r^\theta} = \frac{b}{\theta(1+b)}\left(-  e^{-2r^\theta} + \frac{b-1}{b}e^{-r^\theta} + \frac{1}{b}\right).
\end{equation}
  The r.h.s. is a polynomial 
 function (call it $P_{\theta,\beta}$) of the variable $e^{-r^{\theta}}$. Let us rewrite Eq. (\ref{extremaEq}) as follows (using the parameter $\beta$ instead of $b$):
\begin{equation}\label{varDef}
\begin{split}
 G(x) =& P_{\theta,\beta}(x),\\
{\mathrm{with}}&\;\; x := e^{-r^\theta},\\
 &G(x) := -  x\log( x ),\\
 &P_{\theta,\beta}(x):= -\frac{\beta}{\theta}( x-1)\left( x - 1 + \frac{1}{\beta}\right).
\end{split}
\end{equation}
 The values of the variable $x$ corresponding to physical parameters of the model are in $]0,1[$.
The function $G$ is a positive quantity for any value of the 
 resetting rate
 The values of the variable $x$ corresponding to local extrema of the mean lifetime 
 must be between the two roots of the polynomial $P_{\theta,\beta}$, 
which are $1$ and  the negative quantity $1-\beta^{-1}$.
 For a given index $\theta$, there are extrema of the lifetime $\left\langle  T^{(r,\beta)}\right\rangle$ if the graphs
  of the functions $G$ and $P_{\theta,\beta}$ have intersection points in $]0,1[$.\\

\begin{figure}
\includegraphics[width=18cm]{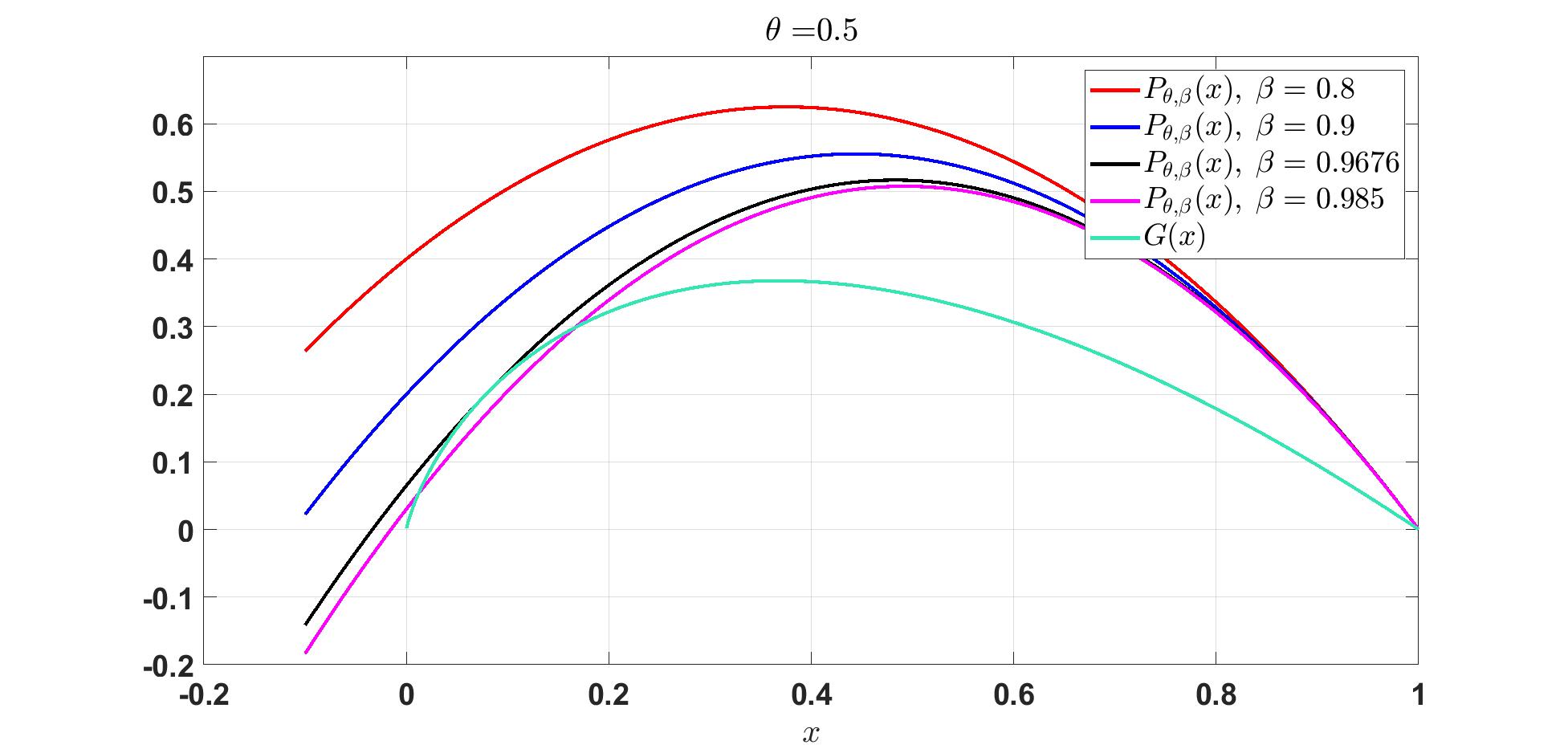}
\caption{The behavior  of the functions $G$ and $P_{\theta,\beta}$ for $\theta = 1/2$. If the sign-keeping probability $\beta$ is greater than a critical value $\beta_c(\theta)$, the graphs of $G$ and $P_{\theta,\beta}$
 intersect at two points. The critical value of $\beta$  in this case is $\beta_c(1/2)\simeq 0.9676$. }
\label{illBehav}
\end{figure}

\begin{figure}
\includegraphics[width=18cm]{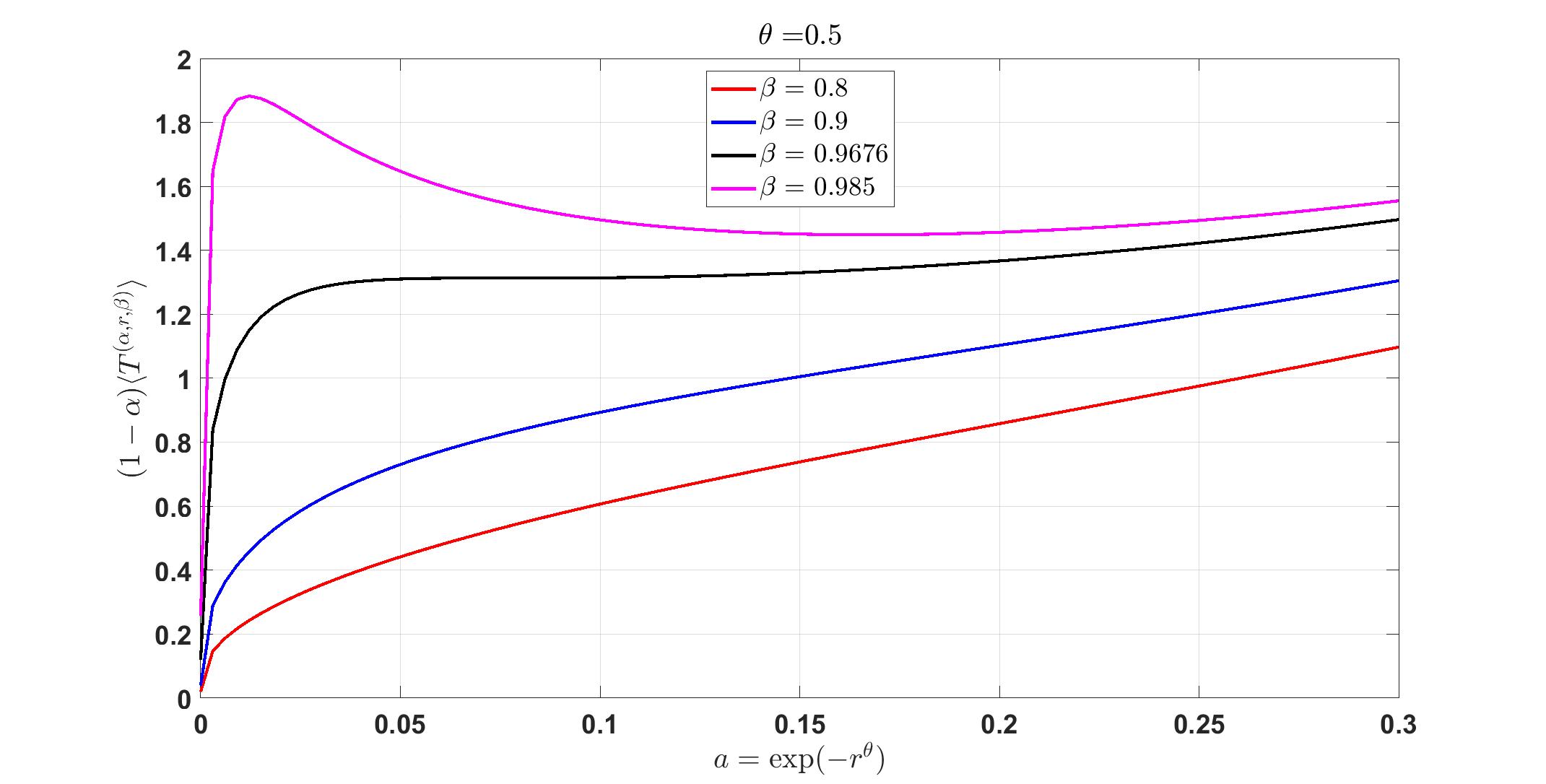}
\caption{{\bf{The average lifetime of the process (up to a factor of $1-\alpha$) as a function of the variable $e^{-r^\theta}$, for $\theta=1/2$}}. The variable takes values in the entire interval $]0,1[$ (for visibility values have been truncated to show the behaviors corresponding to the choice of parameter $\beta$ in Fig. \ref{illBehav}). Two extrema appear when $\beta$ becomes larger than the critical value $\beta_c(1/2)\simeq0.9676$. In the  regime of strong resetting rate, $e^{-r^\theta}$ is close to $0$ and the  average lifetime goes to zero.}
\label{lifeTimeBehavior}
\end{figure}

 At fixed $\theta$, the graphs of the functions $P_{\theta,\beta}$ (for $\beta$ in $]0,1[$)   are a one-parameter family of curves. We need to find the points of intersection between these curves and the graph of $G$. 
  If $\beta$ is low enough, there is no intersection of the two graphs on the relevant interval (i.e. no extremum of the average lifetime). 
 If $\beta$ is close enough to $1$, the 
 negative root of the polynomial function $P_{\theta,\beta}$ is close to $0$, and there are  two  intersection points
 (see  Appendix \ref{functionApp}  for calculations justifying these claims).
 These two regimes are separated by a critical value of the sign-keeping probability, call it $\beta_c(\theta)$, for which the two graphs are tangent to each other.  This behavior is illustrated on Fig. \ref{illBehav} for $\theta = 1/2$. The corresponding lifetimes  are plotted on Fig. \ref{lifeTimeBehavior}.

\subsection{Critical value of the sign-keeping probability}

 Let us look for a parametric expression of the critical value $\beta_c(\theta)$ of the sign-keeping probability.
 Consider some parameter $a$ in $]0,1[$, and let us look for values of the parameters $\theta$ and $\beta = \beta_c(\theta)$ such that the graphs 
 of the functions $G$ and $P_{\theta,\beta_c(\theta)}$ are tangent to each other at the point of coordinates $(a,G(a))$. This condition reads
\begin{equation}
\begin{cases}
 G(a ) &= P_{\theta,\beta_c(\theta)}( a ),\\
 G'(a) &= P_{\theta,\beta_c(\theta)}'(a ),
\end{cases}
\end{equation}
or more explicitly
\begin{equation}\label{tangentoaram}
\begin{cases}
 a \log a &=  \frac{\beta_c(\theta)}{\theta}(a- 1)(a - 1 + \beta_c(\theta)^{-1}),\\
  \log a + 1 &=  \frac{\beta_c(\theta)}{\theta}(2a  - 2 + \beta_c(\theta)^{-1} ).
\end{cases}
\end{equation}
 This is a linear system in of equations in $\beta_c(\theta)/\theta$ and $1/\theta$, which is easily solved (see Appendix \ref{paramApp} for the explicit steps).\\ 

\begin{figure}
\includegraphics[width=18cm]{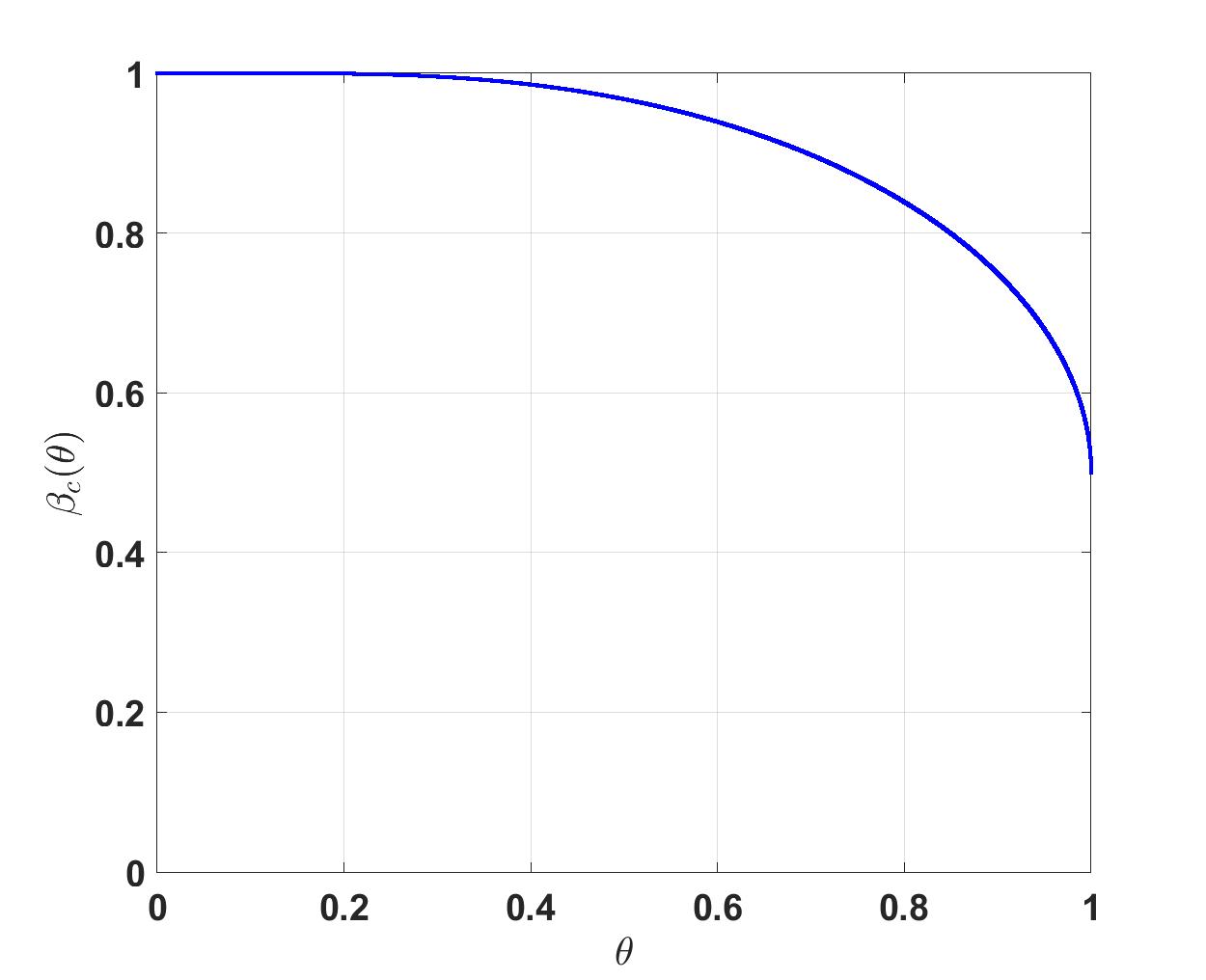}
\caption{{\bf{The critical value of the sign-keeping probability  as a function of $\theta$.}} This graph yields the phase diagram of the model: at a point of coordinates $(\theta,\beta)$ in $]0,1[\times]0,1[$ the mean lifetime $\langle T^{(r,\beta)}\rangle$  is a decreasing function of the resetting rate if $\beta<\beta_c(\theta)$. If $\beta>\beta_c(\theta)$,
    the mean lifetime  has two local extrema in the resetting rate $r$.}
\label{betaThetaParam}
\end{figure}

The resulting parametric equations read 
\begin{equation}\label{betaa}
\begin{split}
  \beta_c(\theta)& =  \frac{ - \log a - 1  + a    }{  - \log a - 1 + 2a + a^2 \log a - a^2    },\\
\end{split}
\end{equation}
\begin{equation}\label{thetaa}
\begin{split}
\theta &=  -\frac{1}{\log a + 1}\left(  \frac{  \log a  + 1  - 2a\log a -2a  + a^2 \log a    + a^2  }{  \log a + 1 - 2a - a^2 \log a + a^2}  \right),
\end{split}
\end{equation}
where we have ordered the terms in each of the factors in decreasing order of dominance in the limit of small $a$ (when $a$ goes to zero, the index $\theta$ goes to $0$ and the critical value $\beta_c(\theta)$ goes to $1$,  see the  next subsection for local properties of the critical parameter $\beta_c(\theta)$ when the index $\theta$ is close to $0$ or $1$).
 Combining these two  parametric equations  allows to plot 
 $\beta_c(\theta)$ against $\theta$ by varying the parameter $a$ in the interval $]0,1[$ (see Fig. \ref{betaThetaParam} ).\\

\begin{figure}
\includegraphics[width=18cm]{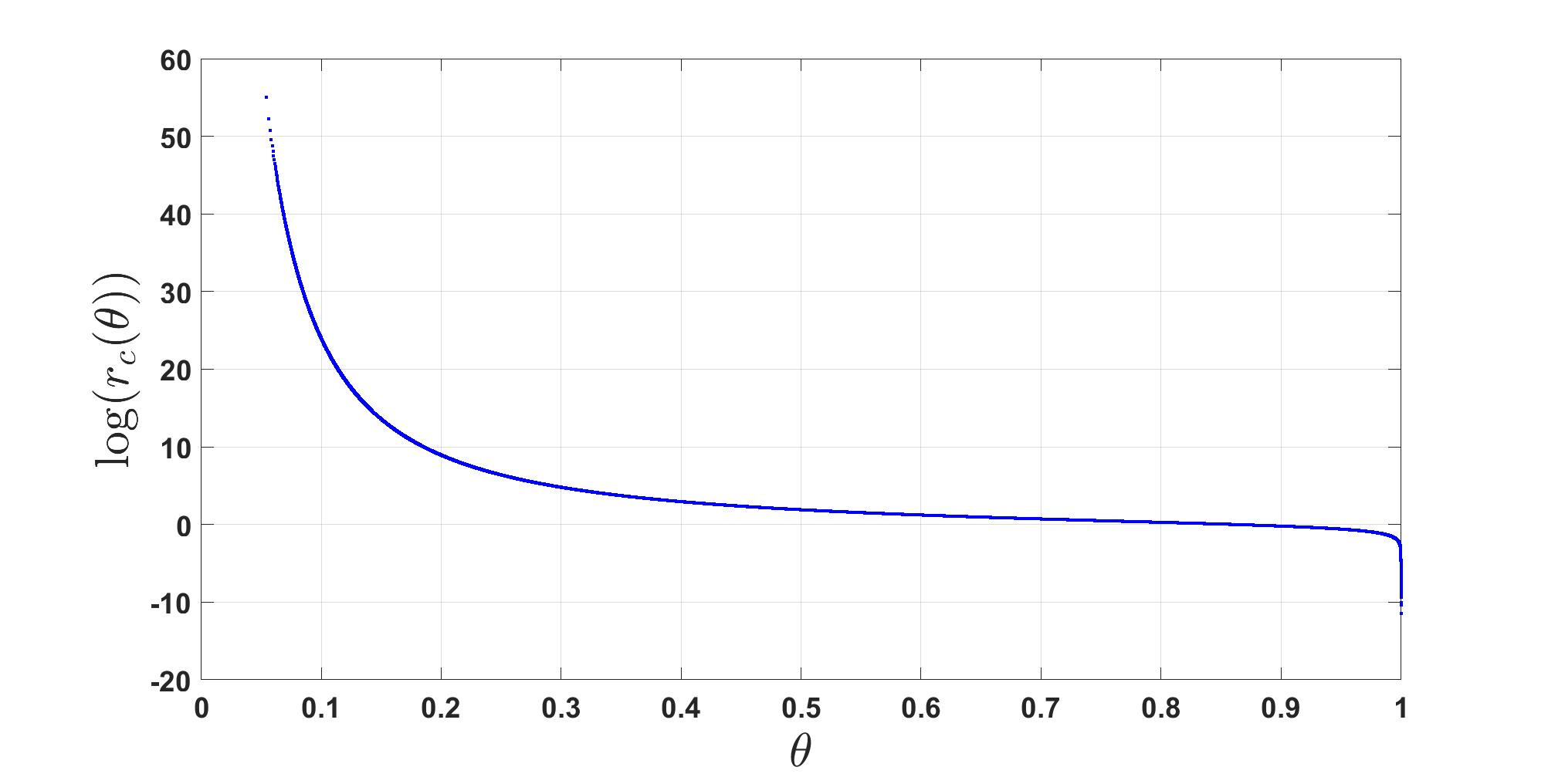}
\caption{{\bf{The logarithm of the resetting rate $r_c(\theta)$ (defined in Eq. (\ref{rca})) corresponding to the critical value $\beta_c(\theta)$.}}  The resetting rate $r_c(\theta)$ goes to infinity when $\theta$ goes to zero, and to $0$ when $\theta$ goes to 1. Equivalents of $r_c(\theta)$ are worked out in Eqs (\ref{rc0},\ref{rc1}).}
\label{rThetaParam}
\end{figure}

 Moreover, the parameter $a$ yields the value $r_c(\theta)$ of the resetting rate at which the lifetime is stationary for $\beta = \beta_c(\theta)$. 
\begin{equation}\label{rca}
r_c(\theta) = \left( -\log( a ) \right)^{\frac{1}{\theta}} =  \left( -\log( a ) \right)^{-\frac{(\log a + 1) (\log a + 1 - 2a - a^2 \log a + a^2)}{ \log a  + 1  - 2a\log a -2a  + a^2 \log a    + a^2  }}. 
\end{equation}
 Varying the parameter $a$ in $]0,1[$ (and using again Eq. (\ref{thetaa}) for the  parametric expression of $\theta$) allows to plot the resetting rate $r_c$ against $\theta$  (see Fig. \ref{rThetaParam}).

\subsection{The limit of low index $\theta$ (or high $\beta_c(\theta)$)}
  
If the process has  a high critical  parameter  (meaning $\beta_c(\theta)$ is close to $1$),
  the position $a$ of the parameter is close to zero, and from Eqs (\ref{betaa},\ref{thetaa}) we obtain
\begin{equation}\label{thetaEq}
\theta \underset{a\to 0}{\sim} -\frac{1}{ \log a},\;\;\;\;{\mathrm{i.e.}}\;\;\;\;\; a \underset{\theta\to 0}{\sim} e^{-\frac{1}{\theta}},
\end{equation}
 and the corresponding value of the index $\theta$ is close to zero.\\

 On the other hand, Eq. (\ref{betaa}) implies that 
 \begin{equation}
 \beta_c(\theta) - 1 \underset{\theta\to 0}{\sim}  \frac{a}{\log a} \sim \theta e^{-\frac{1}{\theta}}.
\end{equation}
 Combining with Eq. (\ref{thetaEq}) yields the behavior of $\beta_c(\theta)$  for low values of  $\theta$:
\begin{equation}
 \beta_c(\theta) - 1 \underset{\theta\to 0}{\sim} \theta e^{-\frac{1}{\theta}}.
\end{equation} 
 
 Substituting the equivalent worked out in Eq. (\ref{thetaEq}) into the expression of the critical resetting rate in Eq. (\ref{rca}) yields
\begin{equation}\label{rc0}
  r_c(\theta) \underset{\theta\to 0}{\sim} \left( \frac{1}{\theta} \right)^{\frac{1}{\theta}}.
\end{equation}

\subsection{The limit of high index  $\theta$}
  If the index $\theta$ is close to $1$, the value of the parameter $a$ is close to  $1$. We can probe the local behavior of 
 $\beta_c$ as a function of $\theta$, close to $\theta = 1$, by expanding the parametric equations in powers of $h$, with 
\begin{equation}
 a = 1-h. 
\end{equation}
From Eq. (\ref{betaa}) we obtain
\begin{equation}\label{weakBeta}
 \beta_c(\theta) = \frac{- \log( 1-h) - 1 + 1 - h}{ -\log(1-h) - 1 + 2( 1-h) + (1-h)^2 \log( 1-h) - (1-h)^2)}. 
\end{equation}
Using the expansion $\log ( 1-h) = - h - \frac{h^2}{2} + o(h^2)$,
 we notice that the terms of order zero and $1$ in $h$ are zero, and that the leading terms 
  are of order $h^2$, both in the numerator and the denominator:
\begin{equation}\label{highTheta}
\beta_c (\theta) = \frac{\frac{h^2}{2} + o(h^2)}{\frac{h^2}{2} + 2  h\times h - \frac{1}{2}h^2 - h^2 +o(h^2)} =  \frac{\frac{1}{2} + o(1)}{1 +o(1)} ,
\end{equation}
hence the critical value $\beta_c$ goes to $1/2$ when the parameter $a$ goes to $1$.\\

 Using Eq. (\ref{thetaa}) we can confirm that $\theta$ goes to $1$ in this limit:
\begin{equation}
 \theta = \frac{2(a-1)\beta + 1}{\log(1-h) + 1} = \frac{1}{1-h+o(h)}\left( 1-\frac{2}{2}h + o(h)\right) = 1 + 0h + o(h).
\end{equation}
 We have therefore derived the limit 
\begin{equation}
 \underset{\theta \to 1}{\lim} \beta_c(\theta) = \frac{1}{2}.
\end{equation} 
Moreover, working out the term of order $h^3$ in the numerator and denominator of the r.h.s. of Eq. (\ref{weakBeta}) yields
 \begin{equation}\label{nextbeta}
 \beta_c(\theta) = \frac{\frac{1}{2}h^2 + \frac{h^3}{3} +o(h^3)}{ h^2 + \left(\frac{1}{3}- 1 \right) h^3 + o(h^3) } = \frac{1}{2} \frac{1 + \frac{2h}{3} + o(h)}{1  -\frac{2h}{3} + o(h)} = \frac{1}{2} \left(   1 + \frac{4}{3} h + o(h) \right),
 \end{equation}
 which implies
 \begin{equation}
\begin{split}
 \theta=& \frac{ 1- \frac{2h}{2}\left(  1 + \frac{4}{3} h + o(h) \right)  }{1- h - \frac{h^2}{2} + o(h^2)}\\
=& \left(  1 -h - \frac{4}{3} h^2 + o(h) \right)\left( 1 + h + \left(\frac{1}{2}+1\right)  h^2 + o(h^2)   \right)\\
=& 1 + 0h + \left(  -1 - \frac{4}{3} +\frac{3}{2}   \right) h^2  + o(h^2)\\
=& 1 - \frac{5}{6}h^2 + o(h^2). 
\end{split}
 \end{equation}
 Combining with Eq. (\ref{nextbeta}) yields
\begin{equation}\label{eqBetaTheta}
 \beta_c(\theta) -\frac{1}{2} \underset{\theta \to 1}{\sim}  \frac{2}{3}  \sqrt{\frac{6}{5}}\sqrt{1- \theta},
\end{equation}
 which indicates that the right-end of the plot in Fig. \ref{betaThetaParam} has a parabolic shape.\\

 On the other hand, the parametric  expression  of the rate $r_c(\theta)$ in Eq. (\ref{rca}) yields
 \begin{equation}
 r_c(\theta) \sim h^{\frac{1}{\theta}}\sim h.
\end{equation}
 Combining with Eqs (\ref{nextbeta},\ref{eqBetaTheta}) we obtain 
\begin{equation}\label{rc1}
  r_c(\theta) \underset{\theta\to 1}{\sim} {\sim} \sqrt{\frac{6}{5}}\sqrt{1- \theta}.
\end{equation}

\subsection{Extrema of the mean lifetime}
 At fixed $\theta$, for $\beta>\beta_c(\theta)$, there is an interval  $[\rmin(\theta,\beta), \rmax(\theta,\beta)]$ on which the mean lifetime 
 of the process is an increasing function of the resetting rate. In the variable $a=e^{-r^\theta}$, this corresponds to an interval 
\begin{equation}\label{noto}
[a_-(\theta,\beta), a_+(\theta,\beta)] = [\exp( -\rmax(\theta,\beta)^\theta ), \exp( - \rmin(\theta,\beta)^\theta) ].
\end{equation}
 
 Let us work out the local behavior of   $a_\pm(\theta,\beta)$ at fixed $\theta$, for $\beta = \beta_c(\theta) + \epsilon$, 
 for some small increment  $\epsilon>0$. 
Let us write the coordinate of an intersection point of the graphs of $G$ and $P_{\theta,\beta}$ as $(a,G(a))$. This point is close to the unique intersection  of the 
 graphs of the functions $G$ and $P_{\theta,\beta_c(\theta)}$, denoted by $(a_c,G(a_c))$, so we write it as
\begin{equation}
 a = a_c + \alpha, 
\end{equation}
for some small $\alpha$. The coordinate $a$  is $a_-(\theta,\beta)$ (resp. $a_+(\theta,\beta)$) if $\alpha<0$ (resp.  $\alpha>0$).\\ 


\begin{figure}
\includegraphics[width=19cm]{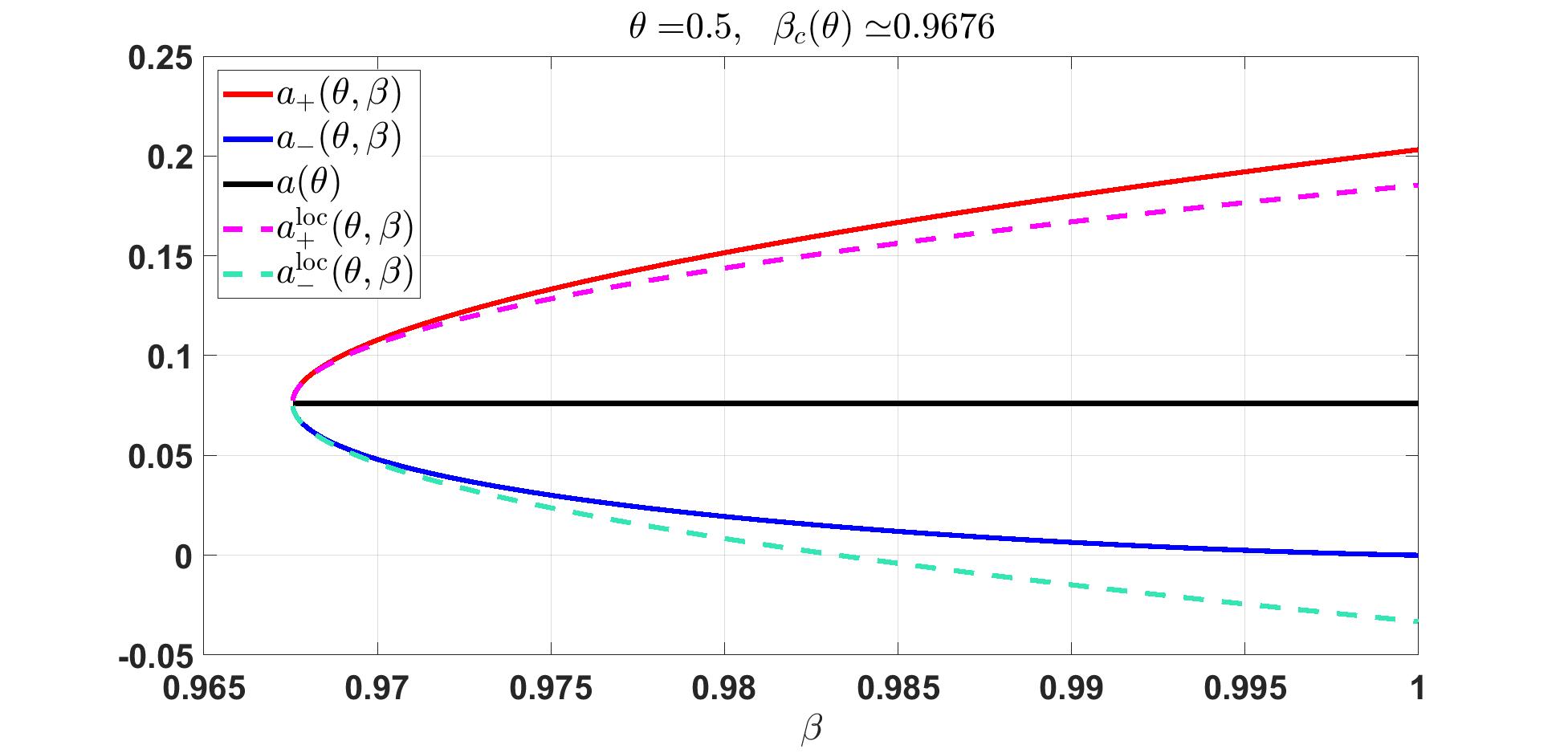}
\caption{{\bf{The  extrema $a_\pm(\theta,\beta)$ as a function of $\beta$ for $\theta = 1/2$.}} The extrema exist for  
 for $\beta>\beta_c(\theta)$. For $\beta$ close to $\beta_c(\theta)$, the local behavior $a^{\mathrm{loc}}_\pm(\theta,\beta)$ predicted by Eq. (\ref{predSqrt}) is shown in dotted lines.}
\label{branchesB}
\end{figure}


The intersection condition reads
\begin{equation}\label{condInter}
 G( a_c + \alpha) = -\frac{\beta_c+\epsilon}{\theta}( a_c + \alpha-1)\left( a_c + \alpha - 1 + \frac{1}{\beta_c+\epsilon}\right).
\end{equation}
 The Taylor expansion of this condition, at order two in $\alpha$ and $\epsilon$, is worked out in Appendix \ref{Taylor}. By definition of the quantity $a_c$, the terms  of order $0$ in $\alpha$ in the above expression sum to zero. This implies that $\alpha^2$ and $\epsilon$ are of the same order. More precisely (see Eq. (\ref{morePrec})),
\begin{equation}\label{predSqrt}
  a_\pm(\theta, \beta_c(\theta) + \epsilon) = a_c(\theta) \pm \frac{1-a_c(\theta)}{\sqrt{\left|\beta_c(\theta) - \frac{\theta}{2 a_c(\theta)}\right|}} \sqrt{\epsilon} + o(\sqrt{\epsilon}).
\end{equation}
 The numerical values of $a_\pm(\theta, \beta)$ for $\beta>\beta_c(\theta)$ are plotted against $\beta$ for $\theta=1/2$ on Fig. (\ref{branchesB}).
 The local estimates of Eq. (\ref{predSqrt}) are shown in dotted lines.\\

 To express the position of the extrema in the resetting rate, using the notations of Eq. (\ref{noto})  as well as the parametric equation (Eq. (\ref{betaa})) for $\beta_c$, we obtain
\begin{equation}
 r_\pm(\theta,\beta_c(\theta) + \epsilon) - r_c(\theta)\underset{\epsilon \to 0^+}{\sim} \pm\frac{1- e^{-r_c(\theta)^\theta}}{\theta r_c(\theta)^{\theta^{-1}} \sqrt{e^{-r_c(\theta)^\theta}\left(  -\frac{\theta}{2} + \beta_c(\theta) e^{-r_c(\theta)^\theta}\right) }}    \sqrt{\epsilon}.
\end{equation}

%
%
%

\section{Discussion}

We have introduced a simple prescription for renewal processes, with a probability $\beta\in]0,1[$ to keeping the sign state at resetting events.
 It can be interpreted in terms of a single spin as follows. We consider a single spin whose sign state persists over a time drawn from the broad distribution 
 $\rho$, as in \cite{baldassarri1999statistics}. We couple it to an external source of energy, which injects a random quantity of energy at Poisson-distributed times. The parameter $\beta$ is the probability that the energy is greater than the fixed energy needed to flip the spin.\\

 Renewal equations for observables of the model can easily be obtained  by conditioning on the latest resetting event. We have illustrated this in detail
 in the case of the survival probability in the presence of a random trap at the origin. We have expressed the mean lifetime of the process in terms  of the 
 parameter $\theta$ and the value $\hat{\rho}(r)$ of the Laplace transform of the density function  $\rho$ of renewal times.  The resetting prescription 
 makes this mean lifetime finite, even for broad distributions of renewal times. The mean lifetime goes to $0$ when the resetting rate goes to infinity (for any density function $\rho$).
 It goes to infinity when the resetting rate goes to $0$ in the case of broad distributions.\\

If renewal times are drawn from a L\'evy stable distribution of index $\theta$, we have found that the behavior of the mean lifetime can be either a decreasing function of the resetting rate $r$, or exhibit two extrema, provided $\beta$ is larger than a critical value $\beta_c(\theta)$. We have not been able to express this critical value of $\beta$ in closed form, but we  have obtained the corresponding 
 phase diagram for $(\theta,\beta)$ in $]0,1[\times]0,1[$ in parametric form. In particular,  there are 
  extrema of the mean lifetime (for some large-enough values of the index $\theta$) if and only if the sign state is kept at resetting with probability $\beta>1/2$. Moreover, the local behaviour of the resetting rates extremizing 
 the mean lifetime, for values of beta close to $\beta_c(\theta)$, is given by the power law $\sqrt{\beta- \beta_c(\theta)}$, with a prefactor dependending on the index $\theta$.\\

 We have disregarded the case $\beta=1$, which corresponds to keeping the current sign state at every resetting event. However, this case is still described 
 by the renewal equation, and the expression of the mean lifetime yields
\begin{equation}
  \left\langle  T^{(\alpha,r,\beta=1)}   \right\rangle\frac{    1-\rhohat(r)   }{   (1-\alpha)r  \rhohat( r ) }= \frac{1}{(1-\alpha)}\frac{e^{r^\theta}-1}{r}.    
\end{equation}
 At large  $r$ the mean lifetime goes to infinity for any value of the index $\theta$ in $]0,1[$, as it is beneficial to reset the position very frequently: resetting events avoid crossing the trap. Moreover, the mean lifetime exhibits a single minimum: in the case $\theta = 1/2$ (up to a factor of  $(1-\alpha)^{-1}$), the mean lifetime is identical to the mean time to absorption of a  one-dimensional diffusive random walker \cite{evans2011optimal} (with an absorbing  trap at unit distance from the origin, unit diffusion constant and resetting to the origin at Poisson-distributed times with rate $r$).\\

The renewal approach we took can be applied to  other observables of renewal processes under stochastic resetting, such as the mean backward recurrence time (the  time elapsed since the latest epoch). Indeed these observables have been characterized in Laplace space in \cite{godreche2001statistics} (even for broad distributions characterized by an index $\theta$ in $]1,2[$). Due to the structure of the renewal equations, the resetting rate is expected to act as a regulator and to make the mean backward recurrence time finite. Technically it becomes difficult 
  to study extrema in the absence of an explicit formula for $\hat{\rho}$.  It would be interesting to generalize the model  to interacting spins, starting with two neighbouring spins whose interaction would be driven by two independent resetting processes switching  their sign states.  \\

\appendix 

\section{Existence of extrema}\label{functionApp}

Consider the following function $H$, whose zeroes in the interval $]0,1[$ correspond to extrema of the mean lifetime in the variable $e^{-r^\theta}$:
\begin{equation}
 H(x) := G(x) -P_{\theta,\beta}(x) = -x \log(x) + \frac{\beta}{\theta}(x-1)\left(x -1 + \frac{1}{\beta} \right).
\end{equation}

Calculating the derivative, 
\begin{equation}\label{derH}
 H'(x) = -\log(x) -1 + \frac{1}{\theta} + \frac{2\beta}{\theta}(x-1).
\end{equation}
 As $-1 + \theta^{-1}>0$ (because $\theta<1$), we can choose $\beta$ small enough   so that the absolute value of the last term satistfies 
\begin{equation}
\left | \frac{2\beta}{\theta}(x-1) \right|< \frac{1}{2}\left(-1 + \frac{1}{\theta}\right). 
\end{equation}
 for all $x$ in $]0,1[$. Hence if $\beta$ is small enough, $H$ is a strictly increasing function on $]0,1[$. Moreover  $H(1)=0$, hence $H$ has no zero on   $]0,1[$ if the sign-keeping probability $\beta$ is small enough.\\

The second derivative reads 
\begin{equation}
 H''(x) = -\frac{1}{x} + \frac{2\beta}{\theta}.
\end{equation}
 It changes sign at $x_c :=\frac{\theta}{2\beta}$. If $\beta$ is close enough to $1$, this value is in the 
 interval $]0,1[$. Substituting into Eq. (\ref{derH}) yields
\begin{equation}
 H'(x_c) = -\log\left( \frac{\theta}{2\beta} \right)+\frac{1-2\beta}{\theta}.
\end{equation}
  When $\beta$ goes to $1$ at fixed $\theta$, this quantity goes to  $-\log\left( \theta/2 \right) - \frac{1}{\theta}$, which is negative (because the function $y\mapsto \log(y/2) - 1/y $ is negative on $]0,1[$). Hence $H'(x_c)<0$ if $\beta$ is large enough. Moreover, $H'(1)=-1 + \theta^{-1}>0$, and $H'(x)$ goes to $+\infty$ when $x$ goes to $0$. Hence $H'$ has two zeroes in $]0,1[$, call them $z_1$ and $z_2$, with $z_1<z_2$. and we have the following variation table:\\

\begin{tikzpicture}
   \tkzTabInit{$x\;\;\;\;$ / 1 , $H'(x)$ / 1, $H(x)$ / 1.5}{$\;\;\;\;\;0$, $z_1$, $z_2$, $1$}
   \tkzTabLine{, +, z, -, z, + }
   \tkzTabVar{-/ $\frac{\beta-1}{\theta}$, +/ $H(z_1)$, -/ $H(z_2)$, +/ 0}
\end{tikzpicture}

The value  $H(0)$ is negative, and $H(z_2)$ is negative. Hence $H$ has  
   two zeroes on $]0,1[$ if $H(z_1)>0$ (which happens if $H(x)$ is positive for some value of  of $x$ in $]0,1[$). 
 When the parameter $\beta$ goes to $1$, at fixed $x$ the quantity 
 $H(x)$ goes to $x(-\log(x)+\theta^{-1}(x-1))$, which is positive for some values of $x$ in $]0,1[$. Hence the value 
 $H(z_1)$ becomes positive when $\beta$ is close enough to $1$. The function  $H$ therefore has exactly two zeroes on the interval $]0,1[$ for large $\beta$.

\section{Parametric equations for the index $\theta$ critical sign-keeping probability $\beta_c(\theta)$}\label{paramApp}

Starting fom the system in Eq. (\ref{tangentoaram}), let us express $\beta_c(\theta)\theta^{-1}$ and $\log a$ in two different ways, using each of the above equations:
\begin{equation}\label{param1}
\begin{cases}
  \log a &=  \frac{\beta_c(\theta)}{\theta}\frac{1}{a}(a- 1)(a - 1 + \beta_c(\theta)^{-1}) = \frac{\beta_c(\theta)}{\theta}(2a  - 2 + \beta_c(\theta)^{-1} ) - 1,\\
  \frac{\beta_c(\theta)}{\theta} &= \frac{  \log a + 1  }{ 2a  - 2 + \beta_c(\theta)^{-1} } = \frac{ a \log a }{(a- 1)(a - 1 + \beta_c(\theta)^{-1})  }.
\end{cases}
\end{equation}
From the second condition in Eq. (\ref{param1}) we obtain the parametric expression 
 of $\beta_c(\theta)$:
\begin{equation}\label{param2}
[ ( a - 1)^2 +  (a-1) \beta_c(\theta)^{-1}] (\log a + 1)= 2a( a-1) \log a + (a \log a) \beta_c(\theta)^{-1},
\end{equation}
 which upon simplification yields the expression of $\beta_c(\theta)$ reported in Eq. (\ref{betaa}). 

Moreover (from the second condition in Eq. \ref{param1}),
 \begin{equation}
\theta (\log a + 1) = 2( a - 1)\beta_c(\theta) + 1, 
\end{equation}
which by substitution yields the parametric expression of $\theta$:
\begin{equation}
\begin{split}
\theta &= \frac{1}{\log a + 1}\left(  2( a - 1)\beta_c(\theta) + 1  \right) =  \frac{1}{\log a + 1}\left(  2( a - 1)\frac{- \log a - 1  + a }{ - \log a - 1 + 2a + a^2 \log a - a^2} + 1  \right)\\
&= -\frac{1}{\log a + 1}\left(  \frac{  \log a  + 1  - 2a\log a -2a  + a^2 \log a    + a^2  }{  \log a + 1 - 2a - a^2 \log a + a^2}  \right),
\end{split}
\end{equation}
  which yields the expression reported in Eq. (\ref{thetaa}).

\section{Local expression of extrema}\label{Taylor}

 The Taylor expansion of Eq. (\ref{condInter}) at order two in the parameters $\alpha$ and $\epsilon$ reads 
\begin{equation}\label{rhstoc}
\begin{split}
 G( a_c ) + G'( a_c) \alpha + \frac{1}{2 a_c}& \alpha^2 +  o(\alpha^2)\\
= &
\left( 1 + \frac{\epsilon}{\beta_c}  \right)(a_c - 1 + \alpha)\left( a_c - 1 + \frac{1}{\beta_c} +\alpha  - \frac{\epsilon}{\beta_c^2} + \frac{\epsilon^2}{\beta_c^3} + o(\epsilon^2)\right),
\end{split}
\end{equation}
 where we have used $G''(x) = x^{-1}$.
 By definition of the quantity $a_c$, the terms  of order $0$ in $\alpha$ in the above expression sum to zero. Expanding on the r.h.s. we obtain
 an expression of the form
\begin{equation}
\frac{1}{2a_c} \alpha^2 + o( \alpha^2) = E\epsilon + M \alpha\epsilon + A\alpha^2 + Q\epsilon^2 + o(\alpha^2,\epsilon^2),
\end{equation}
  for some constant coefficients denoted by $E,M,A,Q$, or
\begin{equation}
 0 = E\epsilon + \gamma \alpha^2 + M \alpha\epsilon + Q\epsilon^2 + o(\alpha^2,\epsilon^2),\;\;\;{\mathrm{with}}\;\;\;\gamma = A - \frac{1}{2 a_c}.
\end{equation}
 The only way to ensure that the terms of order $\alpha$ sum to  zero is for the two leading terms to compensate each other, which implies that 
 $\epsilon$ is of order $\alpha^2$:
\begin{equation}
 0 = E\epsilon + \gamma \alpha^2 + o(\alpha^2).
\end{equation}
 Hence the leading term in the expression of $\alpha$ in terms of $\epsilon$:
\begin{equation}\label{morePrec}
\alpha = \pm\sqrt{\frac{E}{-\gamma}} \sqrt{\epsilon}( 1 + o(1)).
\end{equation}
It is enough to calculate $A$ and $E$ from the product of the three factors on the r.h.s. of Eq. (\ref{rhstoc}).
\begin{equation}
\begin{split}
 A =& \frac{\beta_c}{\theta},\\
 \gamma=& \frac{\beta_c}{\theta}-\frac{1}{2 a_c},\\
  E =& \frac{\beta_c}{\theta}\times \frac{1}{\beta_c}\times (a_c-1) \times( a_c - 1 +\frac{1}{\beta_c})
                        \frac{\beta_c}{\theta}\times (a_c - 1)\times \left(  -\frac{1}{\beta_c^2} \right) = \frac{1}{\theta}( a_c - 1)^2.
\end{split}
\end{equation}
 The values of the parameter $a = a_c + \alpha$ corresponding to the two intersections of   the graphs of $G$ and $P_{\theta,\beta_c(\theta)+\epsilon}$
  are obtained by  substituting these values into Eq. (\ref{morePrec}), leading to the expressions reported in Eq. (\ref{predSqrt}).

\section*{Acknowledgements}
It is a pleasure to thank Linglong Yuan for discussions and correspondence.

\bibliography{bibRefsNew} 

\begin{thebibliography}{10}

\bibitem{evans2011diffusion}
M.~R. Evans and S.~N. Majumdar, ``Diffusion with stochastic resetting,'' {\em
  Physical review letters}, vol.~106, no.~16, p.~160601, 2011.

\bibitem{evans2011optimal}
M.~R. Evans and S.~N. Majumdar, ``Diffusion with optimal resetting,'' {\em
  Journal of Physics A: Mathematical and Theoretical}, vol.~44, no.~43,
  p.~435001, 2011.

\bibitem{gupta2014fluctuating}
S.~Gupta, S.~N. Majumdar, and G.~Schehr, ``Fluctuating interfaces subject to
  stochastic resetting,'' {\em Physical review letters}, vol.~112, no.~22,
  p.~220601, 2014.

\bibitem{evans2018run}
M.~R. Evans and S.~N. Majumdar, ``Run and tumble particle under resetting: a
  renewal approach,'' {\em Journal of Physics A: Mathematical and Theoretical},
  vol.~51, no.~47, p.~475003, 2018.

\bibitem{refractory}
M.~R. Evans and S.~N. Majumdar, ``Effects of refractory period on stochastic
  resetting,'' {\em Journal of Physics A: Mathematical and Theoretical},
  vol.~52, no.~1, p.~01LT01, 2018.

\bibitem{ZRPSS}
P.~Grange, ``Steady states in a non-conserving zero-range process with
  extensive rates as a model for the balance of selection and mutation,'' {\em
  Journal of Physics A: Mathematical and Theoretical}, vol.~52, no.~36,
  p.~365601, 2019.

\bibitem{ZRPResetting}
P.~Grange, ``Non-conserving zero-range processes with extensive rates under
  resetting,'' {\em Journal of Physics Communications}, vol.~4, no.~4,
  p.~045006, 2020.

\bibitem{grange2020entropy}
P.~Grange, ``Entropy barriers and accelerated relaxation under resetting,''
  {\em Journal of Physics A: Mathematical and Theoretical}, 2020.

\bibitem{durang2014statistical}
X.~Durang, M.~Henkel, and H.~Park, ``The statistical mechanics of the
  coagulation--diffusion process with a stochastic reset,'' {\em Journal of
  Physics A: Mathematical and Theoretical}, vol.~47, no.~4, p.~045002, 2014.

\bibitem{magoni2020ising}
M.~Magoni, S.~N. Majumdar, and G.~Schehr, ``Ising model with stochastic
  resetting,'' {\em Physical Review Research}, vol.~2, no.~3, p.~033182, 2020.

\bibitem{grange2021aggregation}
P.~Grange, ``Aggregation with constant kernel under stochastic resetting,''
  {\em Journal of Physics A: Mathematical and Theoretical}, 2021.

\bibitem{grange2020susceptibility}
P.~Grange, ``Susceptibility to disorder of the optimal resetting rate in the
  {L}arkin model of directed polymers,'' {\em Journal of Physics
  Communications}, vol.~4, p.~095018, sep 2020.

\bibitem{perfetto2021designing}
G.~Perfetto, F.~Carollo, M.~Magoni, and I.~Lesanovsky, ``Designing
  nonequilibrium states of quantum matter through stochastic resetting,'' {\em
  Physical Review B}, vol.~104, no.~18, p.~L180302, 2021.

\bibitem{toledo2022first}
J.~Q. Toledo-Marin and D.~Boyer, ``First passage time and information of a
  one-dimensional {B}rownian particle with stochastic resetting to random
  positions,'' {\em arXiv preprint arXiv:2206.14387}, 2022.

\bibitem{grange2022winding}
P.~Grange, ``Winding number of a brownian particle on a ring under stochastic
  resetting,'' {\em Journal of Physics A: Mathematical and Theoretical},
  vol.~55, no.~15, p.~155003, 2022.

\bibitem{sarkar2022synchronization}
M.~Sarkar and S.~Gupta, ``Synchronization in the {K}uramoto model in presence
  of stochastic resetting,'' {\em arXiv preprint arXiv:2203.00339}, 2022.

\bibitem{maso2022conditioned}
A.~Mas{\'o}-Puigdellosas, D.~Campos, and V.~M{\'e}ndez, ``Conditioned backward
  and forward times of diffusion with stochastic resetting: a renewal theory
  approach,'' {\em arXiv preprint arXiv:2205.01613}, 2022.

\bibitem{grangeVoterModel}
P.~Grange, ``Voter model under stochastic resetting,'' {\em arXiv preprint
  arXiv:2207.08590}, 2022.

\bibitem{beysens1986growth}
D.~Beysens and C.~Knobler, ``Growth of breath figures,'' {\em Physical review
  letters}, vol.~57, no.~12, p.~1433, 1986.

\bibitem{bray1994non}
A.~Bray, B.~Derrida, and C.~Godreche, ``Non-trivial algebraic decay in a
  soluble model of coarsening,'' {\em EPL (Europhysics Letters)}, vol.~27,
  no.~3, p.~175, 1994.

\bibitem{derrida1996exact}
B.~Derrida, V.~Hakim, and V.~Pasquier, ``Exact exponent for the number of
  persistent spins in the zero-temperature dynamics of the one-dimensional
  potts model,'' {\em Journal of statistical physics}, vol.~85, no.~5,
  pp.~763--797, 1996.

\bibitem{majumdar1996nontrivial}
S.~N. Majumdar, C.~Sire, A.~J. Bray, and S.~J. Cornell, ``Nontrivial exponent
  for simple diffusion,'' {\em Physical review letters}, vol.~77, no.~14,
  p.~2867, 1996.

\bibitem{baldassarri1999statistics}
A.~Baldassarri, J.~Bouchaud, I.~Dornic, and C.~Godreche, ``Statistics of
  persistent events: An exactly soluble model,'' {\em Physical Review E},
  vol.~59, no.~1, p.~R20, 1999.

\bibitem{godreche2001statistics}
C.~Godr{\`e}che and J.~Luck, ``Statistics of the occupation time of renewal
  processes,'' {\em Journal of Statistical Physics}, vol.~104, no.~3,
  pp.~489--524, 2001.

\bibitem{de2001statistics}
G.~De~Smedt, C.~Godreche, and J.~Luck, ``Statistics of the occupation time for
  a class of gaussian markov processes,'' {\em Journal of Physics A:
  Mathematical and General}, vol.~34, no.~7, p.~1247, 2001.

\bibitem{lamperti}
J.~Lamperti, ``An occupation time theorem for a class of stochastic
  processes,'' {\em Transactions of the American Mathematical Society},
  vol.~88, no.~2, pp.~380--387, 1958.

\bibitem{godreche2015statistics}
C.~Godr{\`e}che, S.~N. Majumdar, and G.~Schehr, ``Statistics of the longest
  interval in renewal processes,'' {\em Journal of Statistical Mechanics:
  Theory and Experiment}, vol.~2015, no.~3, p.~P03014, 2015.

\bibitem{godreche2019non}
C.~Godr{\`e}che, ``Non stationarity of renewal processes with power-law
  tails,'' {\em arXiv preprint arXiv:1909.11540}, 2019.

\bibitem{godreche2023poisson}
C.~Godr{\`e}che, ``Poisson points, resetting and universality,'' {\em arXiv
  preprint arXiv:2302.06536}, 2023.

\bibitem{kusmierz2014first}
L.~Kusmierz, S.~N. Majumdar, S.~Sabhapandit, and G.~Schehr, ``First order
  transition for the optimal search time of {L}{\'e}vy flights with
  resetting,'' {\em Physical review letters}, vol.~113, no.~22, p.~220602,
  2014.

\bibitem{campos2015phase}
D.~Campos and V.~M{\'e}ndez, ``Phase transitions in optimal search times: how
  random walkers should combine resetting and flight scales,'' {\em Physical
  Review E}, vol.~92, no.~6, p.~062115, 2015.

\bibitem{pollard1946representation}
H.~Pollard, ``The representation of $e^{-x^\lambda}$ as a laplace integral,''
  {\em Bulletin of the American Mathematical Society}, vol.~52, no.~10,
  pp.~908--910, 1946.

\bibitem{penson2010exact}
K.~Penson and K.~G{\'o}rska, ``Exact and explicit probability densities for
  one-sided {L}{\'e}vy stable distributions,'' {\em Physical review letters},
  vol.~105, no.~21, p.~210604, 2010.

\end{thebibliography}
\bibliographystyle{ieeetr}

 \end{document}